\documentclass[12pt]{iopart}

\usepackage{iopams}
\usepackage{graphicx}
\usepackage{hyperref}
\usepackage{xcolor}
\usepackage{cite}
\begin{document}

\title[An Electrostatic Lens for ThO Molecules]{Electrostatic focusing of cold and heavy molecules for the ACME electron EDM search}

\author{X. Wu$^{1,2}$, P. Hu$^1$, Z. Han$^1$, D. G. Ang$^2$, C. Meisenhelder$^2$, G. Gabrielse$^3$, J. M. Doyle$^2$, and D. DeMille$^1$}
\address{$^1$ Department of Physics, University of Chicago, Chicago, IL 60637, USA}
\address{$^2$ Department of Physics, Harvard University, Cambridge, MA 02138, USA}
\address{$^3$ Center for Fundamental Physics, Northwestern University, Evanston, IL 60208, USA}
\ead{xinggwu@uchicago.edu}
\vspace{10pt}

\begin{abstract}
The current best upper limit for electron electric dipole moment (EDM), $|d_e|<1.1\times10^{-29}\,$e$\cdot$cm ($90$\% confidence), was set by the ACME collaboration in 2018. The ACME experiment uses a spin-precession measurement in a cold beam of ThO molecules to detect $d_e$. An improvement in statistical uncertainty would be possible with more efficient use of molecules from the cryogenic buffer gas beam source. Here, we demonstrate electrostatic focusing of the ThO beam with a hexapole lens. This results in a factor of $16$ enhancement in the molecular flux detectable downstream, in a beamline similar to that built for the next generation of ACME. We also demonstrate an upgraded rotational cooling scheme that increases the ground state population by $3.5$ times compared to no cooling, consistent with expectations and a factor of $1.4$ larger than previously in ACME. When combined with other demonstrated improvements, we project over an order of magnitude improvement in statistical sensitivity for the next generation ACME electron EDM search. 
\end{abstract}

%
%
%
%
%

\section{\label{sec:level1}Introduction}
The electron electric dipole moment (EDM), $d_e$, is an asymmetric charge distribution along the electron’s spin. Searching for $d_e$ provides a powerful probe for new physics beyond the Standard Model (SM)~\cite{Safronova2018,Chupp2019}. The best upper limit on its value, $|d_e|<1.1\times10^{-29}\,$e$\cdot$cm ($90$\% \textit{c.l.}), was set by the recent ACME measurement~\cite{ACMECollaboration2018}, which we refer to as ACME II. This result severely constrains the parameter space for any new $CP$-violating interactions mediated by particles in the energy range of $3\sim30\,$TeV~\cite{Cesarotti2019}. Further improvement in the electron EDM search would test many theories with new physics at even higher energies.

The ACME experiment is based on measuring spin precession of electrons in thorium monoxide (ThO) molecules, which are present in the form of a cryogenically cooled beam. The EDM search uses the metastable $H\,^3\Delta_1$ state of ThO. This state possesses an effective intra-molecular electric field of $\mathcal{E}\approx78\,$GV/cm~\cite{Skripnikov2016,Denis2016}, which interacts with the electron EDM to produce the sought-for spin precession. To further improve the ACME experiment, several upgrades are underway. 

Envisioned already at the outset of the ACME program\cite{Vutha2010_JPhysB} but not previously demonstrated, is the more efficient use of the molecules produced at the source. Up to now, the ThO beams in ACME have been formed by simply trimming the highly divergent beam with a series of aligned apertures. As a result, only $\sim0.04\%$ of the ThO produced in the initial beam could go through the measurement region, dictated by the solid angle subtended by the detection volume as seen from the source. One solution to increasing the number of useful molecules is to focus the ThO beam with an electromagnetic lens, which can exert a linear restoring force transverse to the beam and direct the originally diverging molecular trajectories into the downstream detection volume. Moreover, the upcoming version of ACME (referred to as ACME III) plans to operate with a longer spin-precession region, designed for an increased coherence time, based on the recently-measured longer lifetime of the $H$ state~\cite{Ang2022}. This upgrade of a longer beamline makes collimation from a molecular lens even more important, since the detection volume is further downstream. Recent spectroscopic studies on ThO~\cite{Wu_2020} indicate that an electrostatic lens could enhance molecular signal by focusing molecules in their long-lived, metastable, and highly polarizable $Q\,^3\Delta_2$ electronic state. To perform the EDM search, molecules would be transferred into the EDM-sensitive $H$ state after the lens, and the subsequent experimental protocol would be the same as in ACME II.

Taming molecular beams via electrostatic interactions has been realized in various forms, including the deceleration and trapping of polar molecules~\cite{Bethlem2000,Wu2017,Aggarwal2021,Prehn2021}. Use of electrostatic lenses has also been applied previously, for the first ammonia maser~\cite{Gordon1955}, in molecular scattering studies~\cite{Kuipers1988,Jongma1997}, and for precision measurements with molecules~\cite{Cho1989,Aggarwal2018,Grasdijk2021}. In this work, we report the electrostatic focusing of a cryogenic buffer gas cooled beam (CBGB) of ThO molecules, which gives rise to a factor of $16$ enhancement in the detectable molecular flux in a test beamline similar to ACME III geometry. The slow mean forward velocity of the beam, $\bar{v}_{f}$, and low temperatures of the translational and rotational degrees of freedom (all standard properties of CBGBs, as shown in various applications~\cite{Hutzler2012,Shaw2020,Gantner2020}), make our lens unusually effective
(see Section\,\ref{subsec:advantage}).

We also implement new and improved protocols for rotational cooling and single quantum state transfer, which maximize the useful signal from beam focusing and the preparation of molecules in the EDM-sensitive $H$ state. The rotational cooling, which occurs upstream from the lens, accumulates molecular population into the absolute ground state $X\,^1\Sigma^+\,|JM\Omega=0,0,0\rangle$ \footnote{$J$ and $M$ follow the usual definition of angular momentum quantum numbers, and $\Omega=\Lambda+\Sigma$, where $\Lambda$ and $\Sigma$ are the projection of total electronic orbital angular momentum $L$ and total electronic spin $S$ on the internuclear axis, respectively.}. We develop a physically compact system that allows a relatively short distance from the molecular source to the lens, hence providing a large effective solid angle of molecules incident on the lens. To prepare molecules for beam focusing, Stimulated Raman adiabatic passage (STIRAP)~\cite{Bergmann1998,Bergmann2019} transfers molecules into $|JM\Omega=2,2,-2\rangle$ in the $Q\,^3\Delta_2$ state, chosen for optimal lens operation. The STIRAP transfer from $X$ to $Q$ in ThO was recently demonstrated~\cite{Wu_2020} to have $90\%$ efficiency for molecules within the entire spatial and velocity capture range of the lens. The combination of electrostatic focusing and these efficient quantum state manipulations makes this molecular lens system a substantial upgrade for ACME III.
  
The content of this paper is as follows. In Section\,\ref{sec:Qstate} we briefly review the properties of the $Q$ state, which enable efficient electrostatic focusing of the ThO beam. Section\,\ref{sec:hexapole} introduces an analytical model of the molecular lens, which captures the principal concepts of lens design and performance. Section\,\ref{sec:setup} describes the experimental setup. Section\,\ref{sec:stateprep} discusses rotational cooling and the STIRAP transfer. Finally, in Section\,\ref{sec:beamfocusing} we present results of electrostatic lensing of ThO in a realistic ACME beamline geometry, and compare with expectations from Monte Carlo trajectory simulations.

\section{\label{sec:Qstate}Molecular state for lensing}
\begin{figure}
	\centering
		\includegraphics[width=0.99\textwidth]{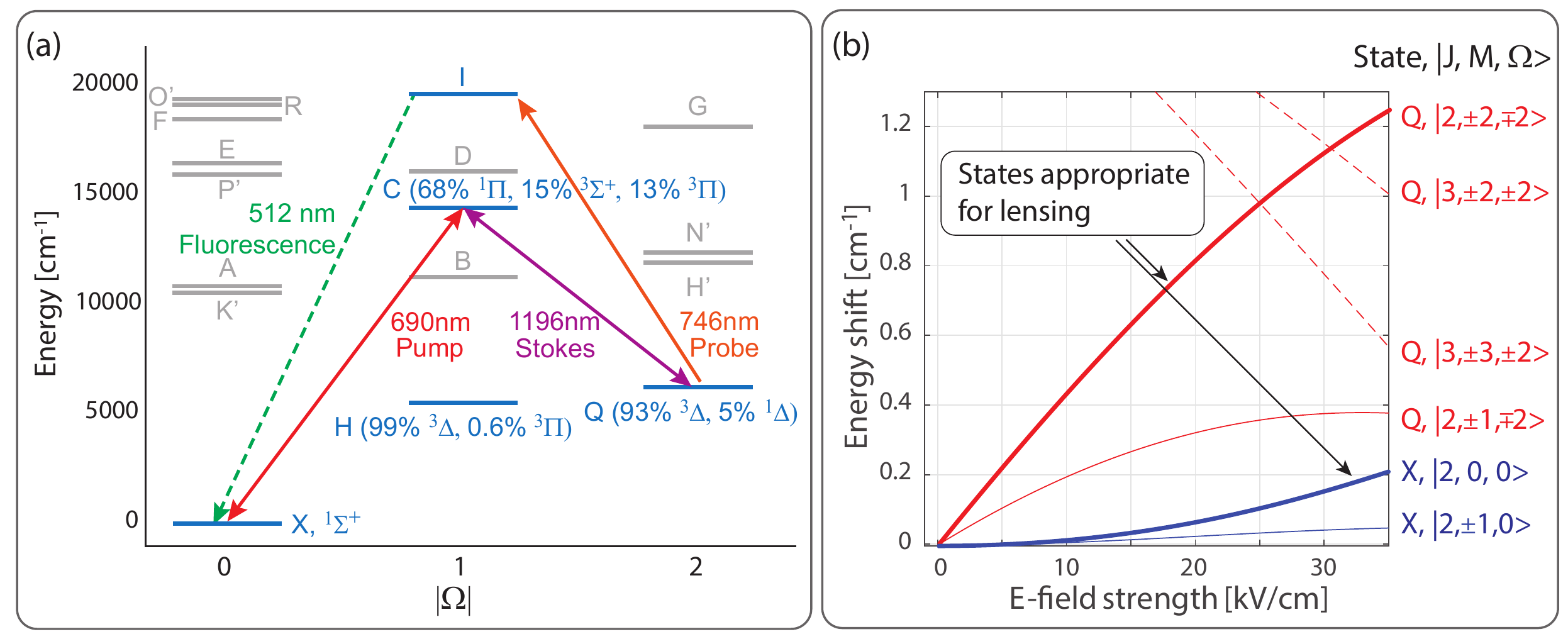}
	\caption{\textbf{(a)} The lowest-energy electronic states of ThO ($v=0$ vibrational states shown). Energy levels (transitions) relevant in this work are shown in blue (with arrows).  Estimated leading contributions from Hund's case (a) terms~\cite{Kaledin2019} are shown where relevant. \textbf{(b)} Stark shift vs. applied electric field strength for selected rotational levels in the $Q$ and $X$ state. All $M\neq 0$ levels are doubly degenerate in an external $\mathcal{E}$-field.  Levels with $M\Omega < 0$ (red solid curves) are low-field seeking. These levels, present in the $Q$ state but not in $X$, exhibit a strong linear Stark shift for small fields; deviation from linearity is substantial at larger fields (when the Stark shift is comparable to the splitting to the nearest rotational level).  All levels with $M\Omega=0$  (hence all $X$ state levels) have weaker, quadratic Stark shifts for small fields. All low-field seeking sublevels of the $J=2$ states are shown for both $X$ and $Q$ states. The zero of energy is set at the field-free $J=2$ state energy for both electronic states. The $J=3$ states (dashed lines) have higher zero-field energy due to rotational splitting.}
	\label{fig:levelscheme}
\end{figure}

Among the known and predicted electronic states of ThO (Fig.\,\ref{fig:levelscheme}(a)), we identified the $Q\,^3\Delta_2$ state to be an ideal candidate for molecular beam focusing. Its relevant properties include~\cite{Wu_2020}:
\begin{itemize} 
\item A long lifetime, $\tau_Q>62\,$ms ($90\%$ \textit{c.l.}), such that negligible loss from decay occurs along the entire ACME beamline, where the total fly-through time is $\approx11\,$ms.
\item A strong optical transition to the $C\,^1\Pi_1$ state of ThO, which enables STIRAP transfer of the population between $X$ and $Q$ via $C$ with about $90\%$ efficiency in both directions.
\item A strong linear Stark shift (Fig.\,\ref{fig:levelscheme}\,(b)), with a measured molecule-frame electric dipole moment, $d_Q\approx4.1\,$D. At the maximum usable field strength (due to onset of nonlinearity in the Stark shift), $\mathcal{E}_{nl}\approx 30\,$kV/cm, the Stark interaction gives rise to a trap depth of $T_{trap}\approx1.6\,$K. This corresponds to a lens capture velocity $v_{cap}=\sqrt{2k_BT_{trap}/m}\approx10\,$m/s for ThO molecules prepared in the low-field-seeking $Q\,|JM\Omega=2,2,-2\rangle$ level (where $k_B$ is the Boltzmann constant and $m$ is the molecular mass).
\end{itemize}

We note briefly the disadvantages that would be encountered in using other long-lived states of ThO for focusing.  For example, the $X\,|JM\Omega=2,0,0\rangle$ ground state has a nearly 10-fold weaker Stark shift than the $Q\,|JM\Omega=2,2,-2\rangle$ state (see Fig.\,\ref{fig:levelscheme}\,(b)), making it far less effective for electrostatic focusing. The $H\,^3\Delta_1$ state has similar Stark shifts to the Q state. However, its much shorter lifetime of $\sim4\,$ms~\cite{Ang2022} would lead to significant loss from decay over the focusing beam path. In addition, dramatically higher laser power is needed to populate the H state via STIRAP~\cite{Panda2016}, for the full distribution of molecules that can be captured by the lens.

\section{\label{sec:hexapole}Analytic model of electrostatic hexapole lens}
\subsection{\label{subsec:principle}General principle}
Electrostatic focusing of molecular beams is closely analogous to focusing beams of light. Hence, widely-known properties of geometric ray optics with thick lenses offer an intuitive guide to designing an electrostatic lens and anticipating its performance~\cite{Cho1991}. The essential goal is to image the molecular beam source at a location near where the molecules will be detected for the EDM search, while achieving the highest possible numerical aperture. This analytical model provides an accurate estimate of the basic lens design parameters, which can then be optimized further using a detailed simulation of molecular trajectories. Here we review the essentials for electrostatic hexapole focusing, appropriate for molecular states with linear Stark shifts. 

\begin{figure}
	\centering
		\includegraphics[width=0.8\textwidth]{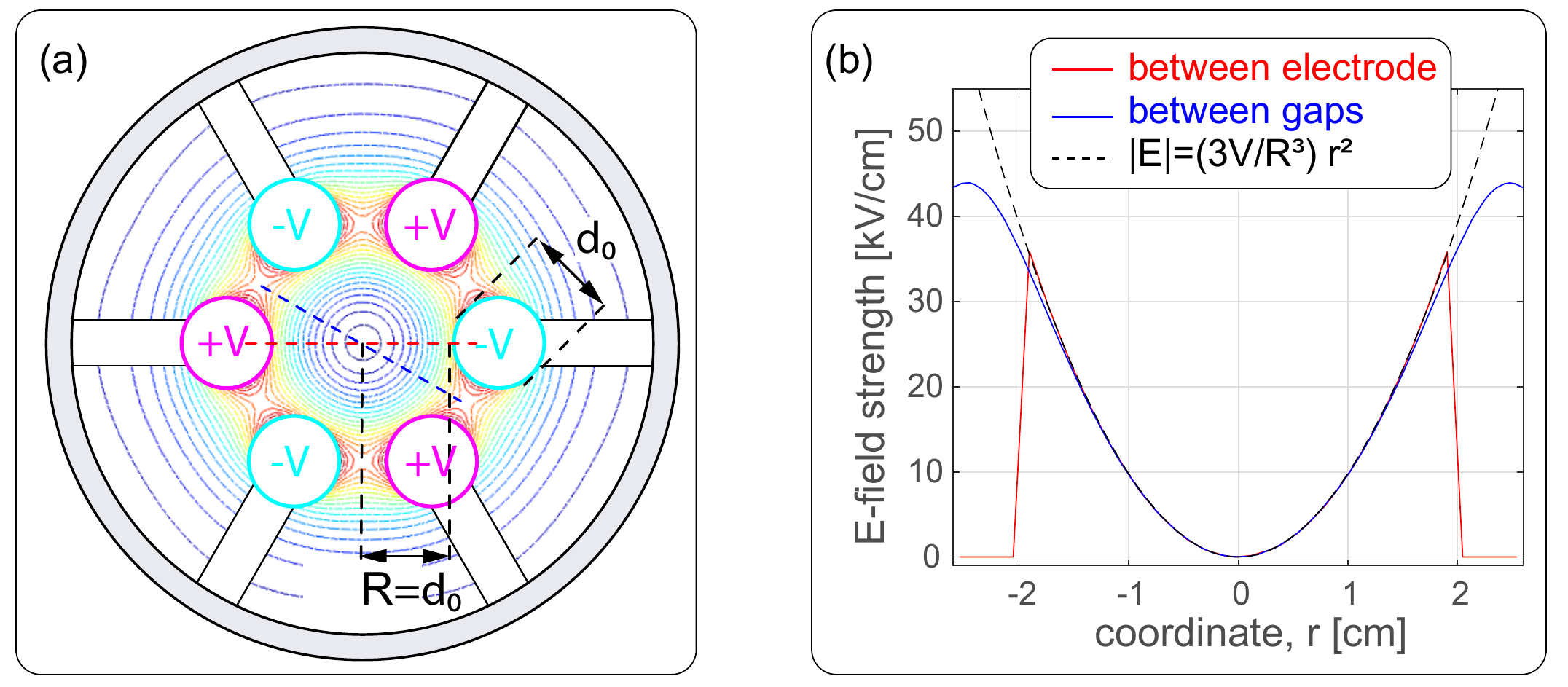}
	\caption{\textbf{(a)} Cross-section of the hexapole lens and its $\mathcal{E}$-field strength contour lines. In our design, the lens radius $R=d_0\approx2\,$cm, where $d_0$ is the electrode diameter, and a typical voltage of $\pm V\lesssim\pm20\,$kV is applied on the electrodes. \textbf{(b)} $\mathcal{E}$-field strength distribution along the two cut-lines in \textbf{(a)}, between two opposite electrodes and between two opposite gaps. Also shown is the analytical model for an ideal hexapole field distribution. The deviations from the ideal distribution here are analogous to astigmatism in optical lenses.}
	\label{fig:hexapole}
\end{figure}

An electrostatic hexapole lens consists of $6$ electrodes arranged at the vertices of a regular hexagon (Fig.\,\ref{fig:hexapole}\,(a)). For technical simplicity we use cylindrical electrodes. A DC voltage $\pm V$ is applied, with alternating polarities on adjacent electrodes. This results in a quadratic field strength distribution near the central axis of the hexapole (Fig.\,\ref{fig:hexapole}\,(b)), given by $\mathcal{E}(r)=(3V/R^3)r^2$, where $r$ is the radial coordinate and $R$ is the radius of the molecular lens (i.e. distance from the center to the nearest electrode edge).  

ThO molecules in the $Q\,|JM\Omega=2,2,-2\rangle$ state that pass through the hexapole field will experience a quadratically varying potential, $W_{st}(r)$, due to the positive linear Stark shift: $W_{st}(r) = -\mathcal{D}\mathcal{E}(r)=-(3\mathcal{D}V/R^3)r^2$, where the state-dependent dipole $\mathcal{D}=d_Q\frac{\Omega M}{J(J+1)}$.
This gives rise to a linear restoring force $\mathbf{F}(\mathbf{r})=-\textrm{d}W_{st}/\textrm{d}r\,\hat{r}=-m\omega^2r\,\hat{r}$, which drives molecules into sinusoidal motion in the radial direction, with angular frequency
\begin{equation}
\omega = \left(\frac{d_Q}{m}\frac{-\Omega M}{J(J+1)}\frac{6V}{R^3}\right)^{1/2}.        
\label{eq:omega}
\end{equation}
We can then write the molecule's radial displacement $r(x)$ at any axial displacement $x$ in the hexapole field, knowing its initial radial displacement $r_i$, radial velocity $\dot{r_i}$, and axial velocity $v_x$:
\begin{equation}
r(x) = r_i\cos(px)+\left(\frac{\dot{r_i}}{v_xp}\right)\sin(px),        
\label{eq:r_x}
\end{equation}
where the lens constant $p\equiv\omega/v_x$~\cite{Cho1991}. The evolution of molecular trajectories from the source to the detector can now be described in terms of optical parameters, analogous to the description of optical rays deflected by a thick lens:
\begin{equation}
\left(\begin{array}{cc}  r \\ \dot{r}/v_x \end{array} \right)
=
\left(\begin{array}{cc} 1 & b' \\ 0 & 1\end{array}\right)
\left(\begin{array}{cc} 1 & 0 \\ -f^{-1} & 1\end{array}\right)
\left(\begin{array}{cc} 1 & a' \\ 0 & 1\end{array}\right)
\left(\begin{array}{cc} r_i \\ \dot{r}_i/v_x\end{array}\right),      
\label{eq:lens_equation}
\end{equation}
where $f=[p\sin(pL)]^{-1}$ is the effective focal length of the hexapole lens, $L$ is the electrode length, and $a'$ ($b'$) is the distance from the object (image) plane to the principal plane near the hexapole lens entrance (exit). The principal planes are at a depth $\delta a=\delta b=[1-\cos(pL)]/[p\sin(pL)]$ into the lens, such that $a' \equiv a+\delta a$ and $b'\equiv b+\delta b$ where $a$ ($b$) is the distance from object (image) to the lens entrance (exit).

\subsection{\label{subsec:boundary}Boundary conditions}
Besides the equations for the molecular trajectories, geometric and practical constraints for the ACME III beamline need to be considered when designing the lens. 
\begin{itemize} 
\item We would like to set the distance from the source to the lens entrance to be as small as possible, $a=a_{min}$, to obtain a maximum numerical aperture. In practice, this is determined by how compact we can make the vacuum system at the source (Sec.\,\ref{subsec:source}), the rotational cooling (Sec.\,\ref{subsec:stateprep1}), and the STIRAP region (Sec.\,\ref{subsec:stateprep2}). For ACME III, we find $a_{min}=0.3\,$m is technically feasible.
\item We would like the lens to capture as many trajectories as possible, hence the maximal field strength, $\mathcal{E}_{max}$, should be roughly $\mathcal{E}_{max}\approx\mathcal{E}_{nl}\approx30\,$kV/cm where the Stark shift starts to become nonlinear.
\item Everything within the transverse capture velocity of the lens, $v_{cap}$, needs to actually make it inside the lens entrance. This sets the minimum lens radius $R>R_{min}$, where $R_{min}\approx(v_{cap}/v_x)\cdot a_{min}+\phi_{ap}/2=1.7\,$cm, where $v_x\approx\bar{v}_f\approx220\,$m/s is the typical mean forward velocity of the ACME molecular beam, and $\phi_{ap}=0.6\,$cm is the diameter of the first aperture after the source.
\item The maximum lens radius $R<R_{max}$ is constrained by the operational voltage we apply while still keeping the maximum field strength near $\mathcal{E}_{max}$. To ensure a robust system, we have found it convenient to keep the long-term operational voltage below $\pm V^*\approx\pm20\,$kV. Thus, $R_{max}=3V^*/\mathcal{E}_{max}\approx2\,$cm.
\item To maximize the molecular flux arriving at the detection volume, we would like the lens to make an image somewhere between the entrance aperture and the detection region of the spin precession measurement (at a distance $b_1$ and $b_2$ away from the lens exit, respectively). Thus, the image-to-lens distance, $b$, is constrained to $b_1\approx 0.6\,\textrm{m}<b<b_2\approx1.6\,$m for the ACME III design. 
\end{itemize}

These boundary conditions determine basic design parameters of the lens, or constrain the ranges where the optimizations of these parameters can be performed. As an initial estimate, we set an object distance $a=0.3\,$m, image distance $b\approx1\,$m, lens radius $R\approx2\,$cm, and electrode voltage $ V=20\,$kV. From Eq.\,\ref{eq:omega}, we get an oscillation frequency $\omega=2\pi\times91\,$Hz and the corresponding lens constant is $p=2.6\,\textrm{m}^{-1}$. Using 
\begin{equation}
\frac{1}{a'}+\frac{1}{b'}=\frac{1}{f},
\label{eq:lensequation}
\end{equation}
we can solve for the electrode length, 
\begin{equation}
L=\frac{1}{p}\,\textrm{cot}^{-1}\left(\frac{abp-1/p}{a+b}\right)=0.46\,\textrm{m},
\label{eq:lenslength}
\end{equation}
and the lens focal length,
\begin{equation}
f=\frac{1}{p\sin(pL)}=\frac{\bar{v}_f}{\omega}\frac{1}{\sin((\omega/\bar{v}_f)L)}=0.41\,\textrm{m}.
\label{eq:focal}
\end{equation}
For comparison, the result from Eqn.\,\ref{eq:lenslength} is close to the optimized electrode length, $L=0.53\,$m, obtained from a full numerical optimization taking into account realistic spatial and velocity distributions at the molecular source, the beamline geometry, and the nonlinearity in the Stark shift of the lensing state (see Fig.\,\ref{fig:levelscheme}\,(b)). The forward velocity spread of the molecular beam (nonlinear Stark shift) leads to the analogue of chromatic (spherical) aberration in optical lenses.

\subsection{\label{subsec:advantage}Advantages of CBGB sources for molecular beam focusing}
Equations~\ref{eq:lensequation} to~\ref{eq:focal} show that the lens performance depends critically on the molecular source:
\begin{itemize} 
\item For a supersonic beam source, which has typically several times higher $\bar{v}_f$, the electrode length and the focal length will increase almost proportionally when the lens voltage is kept constant. To keep the same image-to-lens distance, the lens entrance must be placed further from the source, which reduces the solid angle subtended by the lens quadratically with $\bar{v}_f$.
\item For a thermal source, which has a wide forward velocity spread $\Delta v_f\sim\bar{v}_f$, beam focusing will suffer severely from chromatic aberration, as $f\propto v_f$ approximately. In addition, since the lens frequency $\omega$ depends on the internal state, only a tiny fraction of rotational states populated in a thermal beam will be focused correctly.
\end{itemize}
Consequently, a CBGB molecular source operated in the hydrodynamic regime~\cite{Patterson2007,Hutzler2012} is ideal for electrostatic focusing, due to the small values of $\bar{v}_f$ and $\Delta v_f/\bar{v}_f$, and its low rotational temperature.

\section{\label{sec:setup}Experimental setup}
\subsection{\label{subsec:source}Molecular source}
Characterization of the electrostatic lens is performed using the ACME II beamline~\cite{ACMECollaboration2018}, extended so that the source-to-detection length matches approximately that planned for ACME III. ThO molecules are produced by laser ablation of ThO$_2$ in a cell containing Ne buffer gas at $\sim17K$, and from there extracted in a beam with mean forward velocity $\bar{v}_{f}\approx220\,$m/s, $1\,\sigma$ velocity spread of $\Delta v_f/\bar{v}_f\approx7\%$, and rotational temperature $T_{rot}\sim4\,$K~\cite{Hutzler2011a}. A $690\,$nm laser tuned to the $X\to C$, $P(2)$ line ($P(J)$ denotes the $J\to J’=J-1$ transition) of ThO intersects the molecular beam just outside the buffer-gas cell. The signal from absorption of this laser beam is used to normalize the laser-induced fluorescence (LIF) signal downstream to cancel out slow drifts in ablation yield. The same signal is used as the start time in a time-of-flight (TOF) measurement from which the mean forward velocity of the molecular beam $\bar{v}_{f}$ is extracted.

\begin{figure}
	\centering
		\includegraphics[width=0.99\textwidth]{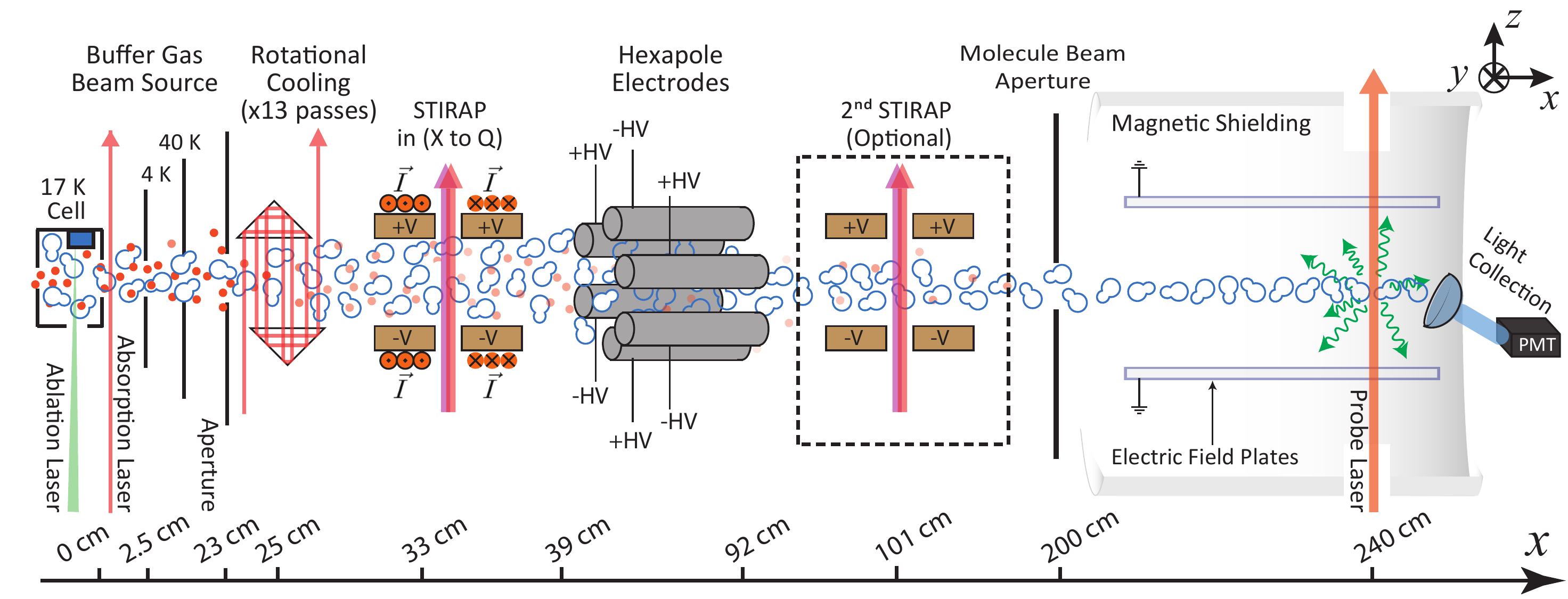}
	\caption{Schematic of the electrostatic hexapole lens test. The coordinate system used throughout this manuscript is in the upper right. The location of each component along the beamline axis is labelled at the bottom. The drawing is not to scale, and emphasis is given to the critical components for the lens test. A $6\,$mm diameter aperture at $x=2.5\,$cm acts as a $4\,$K cryogenic pump for buffer-gas atoms. A $13\,$mm diameter aperture at $x=23\,$cm prevents the direct line of sight for particles from hitting either the rotational cooling optics or the lens electrodes. The distance from the source to the entrance plane of the lens is $a=0.39\,$m, the shortest length possible with the ACME II beam source. Another $2.5\,\textrm{cm}\times2.5\,\textrm{cm}$ aperture at $x=200\,$cm prevents the molecular beam from hitting the electric field plates.}
	\label{fig:setup_scheme}
\end{figure}
\subsection{\label{subsec:stateprep1}State preparation region 1: rotational cooling}
It takes two steps to transfer molecules from the $\sim4\,$K Boltzmann distribution over rotational states in the CBGB into the single, desired lensing state $Q\,|JM\Omega=2,2,-2\rangle$. First, a rotational cooling stage optically pumps molecules into the absolute ground state, $X\,|JM\Omega=0,0,0\rangle$. Second, a STIRAP process transfers the population from $X$ into $Q$. 

The rotational cooling stage consists of a set of multi-pass optics which allows the rotational cooling laser beams to intersect the molecule cloud $13$ times, within a $2.5\,$cm total interaction length. All optics are mounted inside the vacuum chamber to keep the system compact and hence to minimize the undesired expansion of the molecular beam before it reaches the lens entrance. The in-vacuum multi-pass system consists of a pair of anti-reflection coated fused silica (FS) right-angle prisms of $3.6\,$cm hypotenuse-length, offset by $3$mm (see Fig.\,\ref{fig:setup_scheme}), bonded to a FS base plate using UHV compatible and UV-curable epoxy \footnote{Model: \textit{OPTOCAST} 3553-LVUTF-HM. We measured the outgassing rate to be $\lesssim10^{-12}\,\textrm{torr}\cdot$L/s/cm$^2$ with a residual gas analyzer.}. The $13$-pass optical path is first aligned outside the vacuum, with uncured epoxy in place, then exposed to a UV lamp to fix the relative position of each component. The assembly is subsequently heat treated to cure sections that are possibly underexposed to the UV light, and vacuum-baked to $150\,$ degree C before the final installation inside the beamline. The exiting laser beam is directed by a pair of FS mirrors (epoxied on the same base plate) though a vacuum window where the transmitted power can be monitored. A pair of copper field plates is used to apply an $\mathcal{E}$-field of between $0$ and $\approx150\,$V/cm along the $\hat{y}$-axis (perpendicular to both the laser propagation axis, $\hat{z}$, and the beamline axis, $\hat{x}$).

\subsection{\label{subsec:stateprep2}State preparation region 2: STIRAP transfer to the lensing state}
The STIRAP transfer from $X\,|JM\Omega=0,0,0\rangle$ to $Q\,|JM\Omega=2,2,-2\rangle$ needs to avoid populating other $M$ and $\Omega$ sublevels of the $Q\,|J=2\rangle$ manifold. The degeneracy of $\Omega~(M)$ is lifted by applying homogenous fields of $|\mathcal{E}|\approx 50\,$V/cm ($|\mathcal{B}|\approx5\,$G) along the $\hat{z}$-axis, generated by in-vacuum field plates (coils). The $\sigma$-polarized STIRAP laser beams go through slit openings on the $\mathcal{E}$-field plates before intersecting the molecular beam. The level diagram of the STIRAP transfer is detailed in Sec.\,\ref{subsec:stirap}. 

\subsection{\label{subsec:electrodes}Electrostatic lens}
The electrostatic hexapole lens system consists of $6$ stainless steel cylindrical electrodes $53\,$cm in length, $19\,$mm in diameter, and with $9.5\,$mm edge-to-edge spacing between adjacent electrodes (see Fig.\,\ref{fig:hexapole}(a)). Up to $\pm30\,$kV high voltage (HV) is applied, corresponding to an $\mathcal{E}$-field as large as $63\,$kV/cm between adjacent electrodes.\footnote{Multiple Geiger-M\"{u}ller counters are installed around the lens system to monitor possible X-ray generation during HV operation~\cite{West2017}. X-ray count rates above the background level trigger a safety interlock to turn off the HV system.} The exact dimensions of the lens system are optimized for maximal molecular flux at the detection volume downstream, as determined by trajectory simulations that include non-ideal behaviors (\textit{e.g.} molecular beam velocity and spatial spreads, nonlinearity in the Stark shift of the $Q\,|JM\Omega=2,2,-2\rangle$ state, \textit{etc.}). The entire hexapole electrode assembly sits on a home-built $5$-axis kinematic mount, which allows \textit{in situ} in-vacuum alignment. 

\subsection{\label{subsec:stateprepII}State preparation region 3: STIRAP transfer back to the ground state}
To perform the same spin precession measurement protocol as in ACME II, molecules must enter the EDM measurement region in the $X\,|JM\Omega=0,0,0\rangle$ state.  Hence, molecules exiting the lens in $Q\,|JM\Omega=2,2,-2\rangle$ should be transferred back to the ground state. This can be accomplished via $Q\textendash C \textendash X$ STIRAP, performed just after the end of the lens electrodes (similar in design to the pre-lens STIRAP region). We note that $90\%$ efficient $Q\textendash C \textendash X$ STIRAP transfer was demonstrated in~\cite{Wu_2020}. This paper focuses on the performance of the state preparation and electrostatic focusing, so for all of the results described here, this post-lens STIRAP region is not used and is marked as optional in Fig.\,\ref{fig:setup_scheme}. Instead, we monitor only signals from molecules exiting the lens in the $Q$ state. 

\subsection{\label{subsec:detection}Detection region}
The molecules travel $\approx150\,$cm from the end of the lens to the downstream detection region, where molecular population in the $Q$ state is measured by exciting the $Q\to I$ line at $746\,$nm (see Fig.\,\ref{fig:levelscheme}\,(a)) and detecting the $512\,$nm $I\rightsquigarrow X$ LIF signal. Up to $500\,$mW of $746\,$nm laser light is used in detection, expanded vertically (along the $\hat{y}$-axis) to $\approx3\,$cm $1/e^2$-diameter, and collimated along the $\hat{x}$-axis to about $2\,$mm $1/e^2$-diameter.

\section{\label{sec:stateprep}State preparation before the lens: principle and results}

\subsection{\label{subsec:rotcool}Rotational Cooling}
The rotational cooling scheme is designed to occupy minimal length along the molecular beam and to be capable of saturating molecules in the wide transverse spatial and velocity capture range of the lens. The former minimizes the transverse size of the beam that contains molecules within the lens capture range. The latter ensures the entire population from the lens capture range can be accumulated in the ground state.

\begin{figure}
	\centering
		\includegraphics[width=0.95\textwidth]{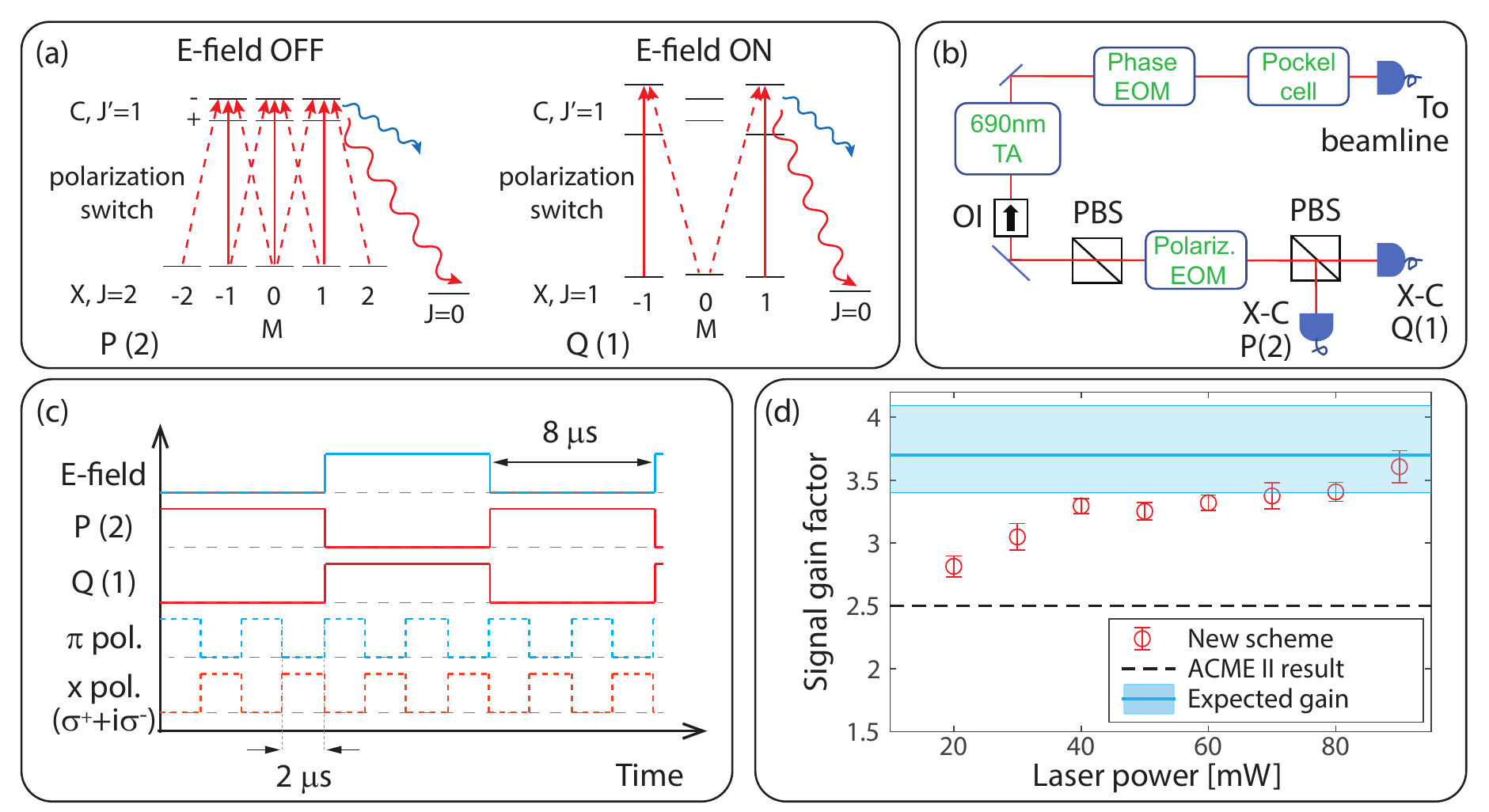}
	\caption{Rotational cooling scheme and result. \textbf{(a)} Level diagrams for rotational cooling on the $X\to C$, $P(2)$ and $Q(1)$ lines of ThO. The straight arrows depict the rotational pumping lines. The squiggle arrows depict spontaneous decays, into the desired final state (red), and out of the optical pumping system (blue). \textbf{(b)} Schematic of the optical components to combine, amplify, and modulate the two laser beams driving the rotational cooling transitions. \textbf{(c)} Modulation timing diagram of the polarizations, frequencies, and $\mathcal{E}$-field applied during rotational cooling. \textbf{(d)} Power dependence of the ground state population gain resulting from rotational cooling. Measurements (red circles) are compared with the expected rotational cooling gain (blue band) and the result obtained in ACME II (black dashed line)~\cite{ACMECollaboration2018}. The blue area indicates the $\pm1\,\sigma$ uncertainty range in the expected gain (see \ref{sec:RotCoolLimit}). Error bars of the data points represent the standard error in the mean over $10$ consecutive measurements, each consisting of $64$ molecule pulses with rotational cooling on and then another $64$ with rotational cooling off.}
	\label{fig:rot_cool}
\end{figure}

Two optical transitions at $690\,$nm are driven for rotational cooling: the $P(2)$ and $Q(1)$ lines ($Q(J)$ denotes the $J\to J’=J$ transition) of the $X\to C$ transition (Fig.\,\ref{fig:rot_cool}\,(a)). These serve to pump population initially in $J=2$ and $1$ levels, respectively, into $J=0$ in $X$. The maximal pumping efficiency on the $P(2)$ line is obtained at zero ambient $\mathcal{E}$-field, under which the parity ($\mathcal{P}$) of the Lambda-doublet in the excited $C\,|J'=1\rangle$ state remains a good quantum number. Population excited to $C\,|J'^{\mathcal{P}'}=1^-\rangle$ (the upper level of the doublet) can only decay to the target state $X\,|J^\mathcal{P}=0^+\rangle$ or back to the initial state $X\,|J^\mathcal{P}=2^+\rangle$. Pumping on the $Q(1)$ line, on the other hand, requires an external $\mathcal{E}$-field to mix the parity eigenstates in $C$. Otherwise, direct optical pumping from $X\,|J^\mathcal{P}=1^-\rangle$ to $X\,|J^\mathcal{P}=0^+\rangle$ is parity-forbidden.

When the $J=2$ and $J=1$ populations are fully depleted by this scheme, the ground state population should increase by a factor of $3.7^{+0.4}_{-0.3}$, 
based on the previously measured vibronic branching ratios from $C$ to all other states~\cite{Wu_2020} and the H\"onl-London factors for the $X\to C$ transition (see \ref{sec:RotCoolLimit}). 
By adding another laser to pump the $X\,|J=3\rangle$ population (using the $P(3)$ line), in principle another factor of $1.17$ increase in population could be gained, but we do not include this extra step here.

To minimize the length needed for rotational cooling, the two laser beams driving the $P(2)$ and $Q(1)$ lines are overlapped in space, then delivered via a common optical fiber (Fig.\,\ref{fig:rot_cool}\,(b)) to the multi-pass optics (see Sec.\,\ref{subsec:stateprep1}) inside the beamline. For molecules to interact with both laser frequencies, we switch rapidly between the two beams (Fig.\,\ref{fig:rot_cool}\,(c)). The $P(2)$ and $Q(1)$ beams are prepared in orthogonal linear polarizations, then combined on a polarization beam splitter (PBS), before passing through a polarization electro-optical modulator (EOM) and an output PBS. We switch the EOM to allow either the $P(2)$ or the $Q(1)$ beam to pass through the output PBS (at typically $15\,$dB extinction ratio) at any given time. For optimal operation, the full frequency-switching cycle is performed in $16\,\mu$s.  
The optical power of this overlapped laser beam is amplified from $10\,$mW to $250\,$mW by a $690\,$nm semiconductor tapered amplifier. Synchronized with the laser beam switching, a homogeneous $\mathcal{E}$-field is also switched between $0$ and $\approx 150\,$V/cm in the region where the laser intersects the molecular beam. To cover the $20\,$MHz FWHM (full width at half maximum) transverse Doppler width of the molecule cloud, the optical beam is spectrally broadened by a phase EOM, adding up to $90$ sidebands of $330\,$kHz spacing. (This sideband spacing corresponds to the natural linewidth of $X\to C$ transition, where the $C$ state lifetime is $\tau_C\approx480\,$ns~\cite{HessThesis2014,Kokkin2014}.) To access all $M$ sublevels in the $J=2$ and $J=1$ levels and to eliminate Zeeman dark states, the polarization of the laser beam is switched between $\hat{s}$ and $\hat{p}$ by a Pockels cell before it is delivered to the beamline. Each polarization is on for $2\,\mu$s, corresponding to about $4\times\tau_C$; this is found empirically to give optimal performance. The rise and fall times of both the frequency and polarization switching pulses are $<20\,$ns, ensuring a sudden (non-adiabatic) change between excitation conditions, as needed for efficient optical pumping in this system. 
The temporal modulation scheme for rotational cooling is summarized in Fig.\,\ref{fig:rot_cool}\,(c). With these, we reduce the length of the rotational cooling region from $\approx20\,$cm in ACME II to $\approx4\,$cm now.

Results when using this compact rotational cooling setup are shown in  Fig.\,\ref{fig:rot_cool}\,(d). We define the gain factor as the ratio of signals from the $X\,|JM\Omega=0,0,0\rangle$ population with the rotational cooling turned on versus off. The data demonstrates that the rotational cooling gain saturates at $\approx3.5$, consistent with expectations based on known branching ratios for decay from the $C$ state~\cite{Wu_2020}. This is a factor of $1.4$ improvement from the ACME II rotational cooling gain~\cite{ACMECollaboration2018}, likely due to the fast  polarization and frequency switching, better $\mathcal{E}$-field control, and higher laser intensity available.

\subsection{\label{subsec:stirap}State transfer via STIRAP}

\begin{figure}
	\centering
		\includegraphics[width=0.55\textwidth]{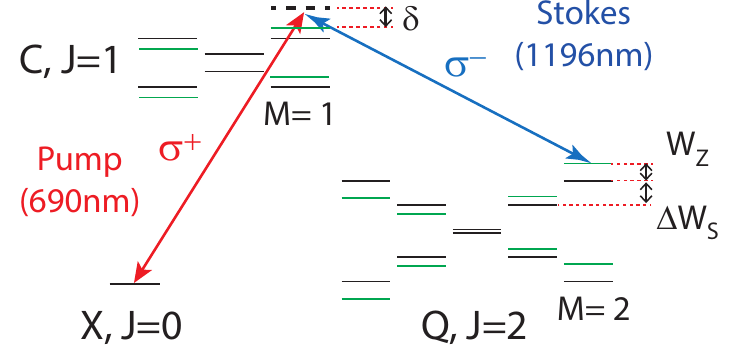}
	\caption{Level diagrams for the STIRAP transfer from $X\,|J=0\rangle$ to $Q\,|JM\Omega=2,2,-2\rangle$ via the intermediate $C\,|JM\Omega=1,1,-1\rangle$ level. Black line segments indicate the Stark-shifted energy levels, and green line segments indicate the Zeeman shifts of each level. The splitting $W_Z$ is the Zeeman shift of $Q\,|JM\Omega=2,2,-2\rangle$, while $\Delta W_S$ is the differential Stark-shift relative to the adjacent $|M|=1$ levels. $\delta$ is the one-photon detuning of the STIRAP laser beams.}
	\label{fig:stirap_scheme}
\end{figure}

The transfer of ThO population from its ground state to the $Q\,|JM\Omega=2,2,-2\rangle$ state used for lensing is performed with near unit efficiency using a stimulated Raman adiabatic passage (STIRAP) process. In our previous studies~\cite{Wu_2020}, a similar STIRAP transfer between $X\,|JM=0,0\rangle$ and $Q\,|JM=2,0\rangle$, via an intermediate state $C\,|JM=1,0\rangle$, was fully characterized, with $90\%$ transfer efficiency  in either direction (\textit{i.e.} $X\textendash C\textendash Q$ and $Q\textendash C \textendash X$). Here, we need to transfer into the stretched state $M=2$ instead of $M=0$. This is realized by applying $\sigma$-polarization in both the Pump ($690\,$nm) and Stokes ($1196\,$nm) beams for the $X\textendash C\textendash Q$ STIRAP (Fig.\,\ref{fig:stirap_scheme}). Since $\pm M$ states are degenerate with an offset $\mathcal{E}$-field alone, to lift the degeneracy between $M=\pm2$ states, and hence prevent STIRAP-transferring to the opposite stretched state due to imperfection in laser polarizations, an offset $\mathcal{B}$-field of $5\,$Gauss is applied in addition to the offset $\mathcal{E}$-field of $50.1\,$V/cm. Both $\mathcal{B}$- and $\mathcal{E}$-fields are along the $\hat{z}$-axis. The former induces a Zeeman shift of $W_Z = 9.6\,$MHz between $M=2$ and $M=0$ levels, while the latter induces a differential Stark-shift of $\Delta W_S=34.2\,$MHz between$|M|=2$ and $|M|=1$ levels. To avoid losses due to spontaneous emission from the $C$ state, the one-photon detuning of the STIRAP laser is blue-shifted by $\delta\approx 18\,$MHz from resonance. Simulations based on prior results show that this STIRAP transfer should saturate over $97\%$ of the population within the lens capture range.

To avoid population loss from $Q\,|JM\Omega=2,2,-2\rangle$ due to spin flips as molecules travel through the volume between the STIRAP region to the hexapole lens, a sufficiently strong $\mathcal{E}$-field needs to be present along the trajectories of molecules. This is achieved using an arrangement of planar electrodes where each STIRAP field plate is aligned with two electrodes in the hexapole lens that have the same polarity (Fig.\,\ref{fig:E_config}). This ensures no reversal of electric field direction along the molecules' trajectories during the transit. Details of the electrode arrangement and the resulting field strength distribution in the transit volume are elaborated in \ref{sec:Efield}. 

\section{\label{sec:beamfocusing}Results from molecular beam focusing}

The electric hexapole lens is optimized for maximal molecular flux at the new ACME detection region (approximately $1.5\,$m downstream from the lens exit). With ThO population prepared in the $Q\,|JM\Omega=2,2,-2\rangle$ state, we characterize its performance and compare it with trajectory simulations. Three key figures are benchmarked: the enhancement factor in the molecular flux, the transverse velocity distribution, and the transverse spatial profile of the lens-focused molecular beam in the detection region. The first of these is an essential factor in projecting the total EDM sensitivity improvement, while the other two determine the laser power requirements for preparation and readout of the spin precession in the ACME III experiment.

\subsection{\label{subsec:totalflux}Molecular flux}

\begin{figure}
	\centering
		\includegraphics[width=0.99\textwidth]{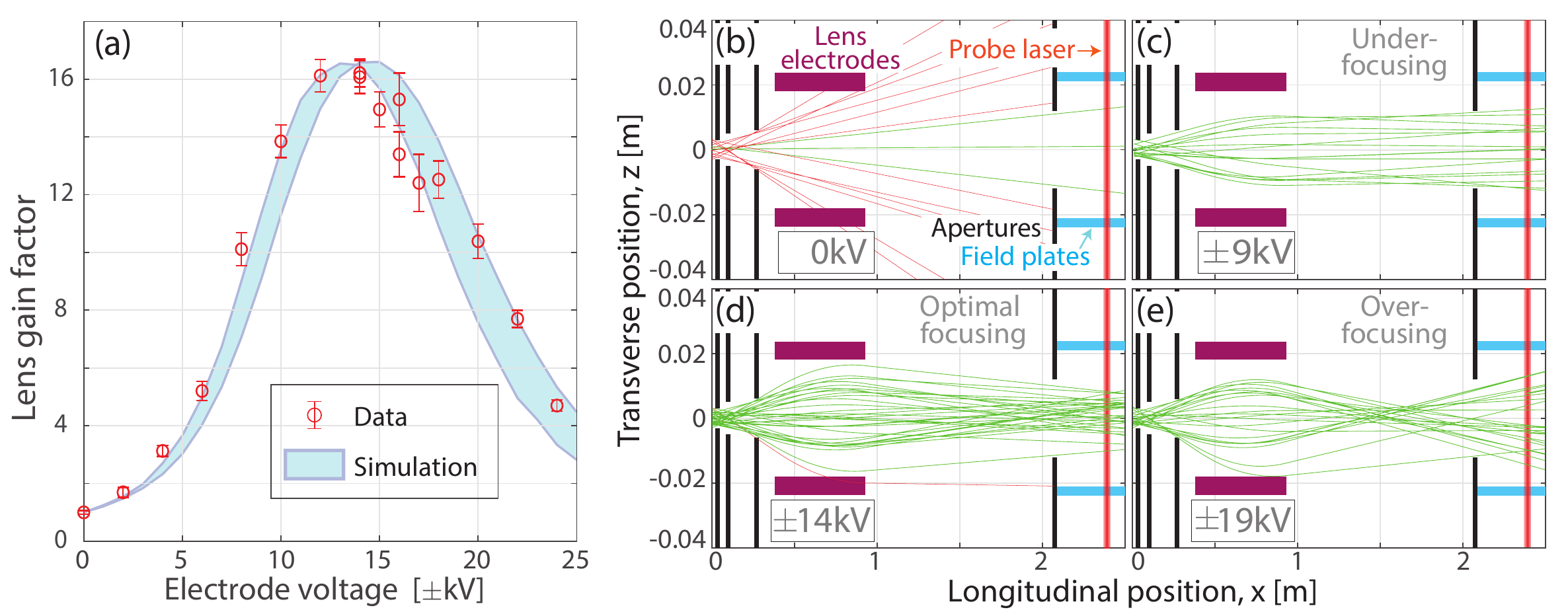}
	\caption{\textbf{(a)} Measured and simulated flux gain from molecular beam focusing versus the voltage applied on the lens electrodes. The lens gain factor is defined as the ratio of signals detected downstream at a given lens voltage versus that with the lens turned off. Error bars of the data points are derived from the standard error in the mean over $\approx6$ consecutive measurements, each consisting of $64$ molecule pulses with lens voltages applied, and then with lens electrodes grounded. The results from trajectory simulations are shown for a range of beam mean forward velocities, $\bar{v}_f$, from $210$ to $220\,$m/s. \textbf{(b)} to \textbf{(e)} Simulated molecular trajectories at selected lens voltages, projected onto the $x$-$z$ plane. Trajectories that make it into the detection region are depicted by green traces, while those with transverse velocity within lens capture range but do not make it to the detection are depicted in red. Black vertical line segments indicate fixed apertures for the molecular beam.}
	\label{fig:integrated_signal}
\end{figure}

The molecular flux enhancement from electrostatic focusing is given by the ratio between the detected molecular signal in the downstream region, with high voltage applied to the lens electrodes versus with the electrodes grounded. This molecular signal is taken at the maximum probe laser intensity, which saturates over $99\%$ of the Doppler spread of the $Q\to I$ transition, and after integrating over the temporal profile of each molecular pulse. The lens gain factor is plotted in Fig.\,\ref{fig:integrated_signal}\,(a), as a function of electrode voltage. As shown there, the largest molecular flux is obtained with $\pm 14\,$kV applied on the hexapole electrodes, yielding an enhancement factor of $16.2(6)$. 

The dependence of the measured enhancement factor with respect to the voltage applied on the lens is consistent with expectations based on our trajectory simulations. The simulation takes into account the realistic spatial distribution of the molecular beam, its transverse and forward velocity spreads, and constraints in beamline geometry. The simulation result, which includes no fitting parameters, is shown in Fig.\,\ref{fig:integrated_signal}\,(a). The good agreement with experimental data is obtained when the exact nonlinearity of the lensing state's Stark shift is included in the simulation~\cite{Wu2016}. The range of simulated results is set by the uncertainty in the molecular beam mean forward velocity $\bar{v}_{f}$, as determined from the \textit{in situ} TOF measurement between the absorption signal upstream and the LIF detection downstream. We observe a small difference in the TOF signal between when the lens is operated at the optimal signal (at $\pm14\,$kV), corresponding to $\bar{v}_{f}=220\,$m/s, and when the lens is turned off, corresponding to $\bar{v}_{f}=210\,$m/s. This $10\,$m/s difference is taken as the uncertainty in $\bar{v}_f$. 

To make sense of the dependence of the gain factor versus applied lens voltage, we take a closer look at the simulated trajectories at selected voltages on the lens, starting from the same initial sample distributions. They are shown in Fig.\,\ref{fig:integrated_signal}\,(b) to (e). The number of green traces shown in each plot is proportional to the predicted integrated signal at the corresponding voltage. For comparison, Fig.\,\ref{fig:integrated_signal}\,(b) and (d) also plot trajectories (at reduced line density for clarity) that are within the lens capture range at $\pm14\,$kV, but do not make it into the detection region. At $0\,$kV (Fig.\,\ref{fig:integrated_signal}\,(b)), the molecules follow ballistic trajectories, since no focusing force is applied. The least number of molecules arrive at the detection region here, due to the beam divergence. As the lens voltage is increased, more trajectories are bent towards the forward direction, thanks to the increasing focusing force. At $\pm9\,$kV (Fig.\,\ref{fig:integrated_signal}\,(c)), the trajectories become nearly collimated after leaving the molecular lens, but the number of trajectories arriving at the detection region is not yet optimized. At $\pm14\,$kV (Fig.\,\ref{fig:integrated_signal}\,(d)), the molecular lens is making an image of the beam source right in front of the molecular beam aperture (at the entrance of the spin precession region, see Fig.\,\ref{fig:setup_scheme}), according to Eqn.~\ref{eq:lensequation}, but blurred by aberrations. This yields the maximal detected signal, which corresponds to about $38\%$ of the transversely slow trajectories (those within the velocity capture range of the lens) out of the beam source, and about $95\%$ of those transversely slow trajectories arriving at the molecular lens. Further increasing the lens voltage causes over-focusing. At $\pm19\,$kV (Fig.\,\ref{fig:integrated_signal}\,(e)), the image plane shifts visibly upstream, to about $60\,$cm before the molecular beam aperture according to Eqn.~\ref{eq:lensequation}. Due to clipping of the over-focused trajectories on this aperture, the detected signal decreases to $\sim70\%$ of the optimal value. 

\subsection{\label{subsec:doppler}Transverse velocity distribution}
\begin{figure}
	\centering
		\includegraphics[width=0.67\textwidth]{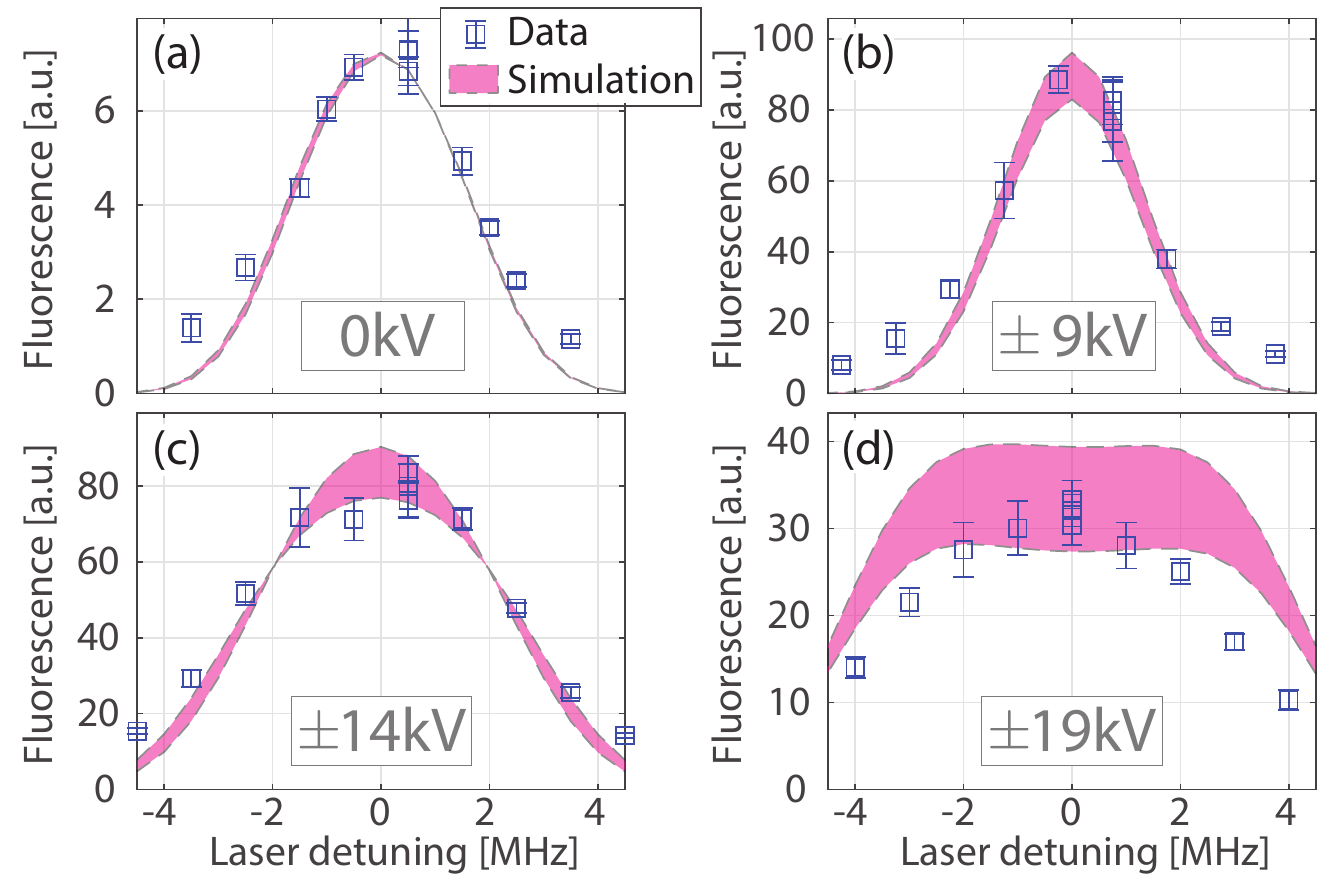}
	\caption{Measured transverse Doppler spectra of the molecular beam in the detection region, at selected lens voltages. Each spectrum is proportional to the transverse ($\hat{z}$) velocity distribution of the beam, where $1\,$MHz detuning corresponds to $0.75\,$m/s velocity. Error bars of the data points represent the standard error in the mean over $\approx4$ consecutive measurements, each consisting of $64$ molecule pulses. The range of predictions from simulated trajectories is set by the uncertainty in $\bar{v}_f$ of the molecular beam, as in Fig.\,\ref{fig:integrated_signal}\,(a).}
	\label{fig:dopplerscan}
\end{figure}
To quantitatively confirm more details of the molecular trajectories, the frequency of the probe laser is scanned to determine the distribution of molecular transverse velocities, via the Doppler shift, at various lens voltages. These Doppler spectra are then compared with expectations based on the trajectory simulations. From the trajectory plots in Fig.\,\ref{fig:integrated_signal}\,(b) and (c), we anticipate that the velocity distribution should narrow when the lens voltage increases from $0$ to $\pm9\,$kV, as the molecular beam gets more collimated and molecular trajectories becomes almost parallel. When the voltage increases further to $\pm14$ and $\pm19\,$kV, we expect the distributions to widen again, as the image plane moves upstream of the detection region (Fig.\,\ref{fig:integrated_signal}\,(d) to (e)).

The experimental results confirm both the expected lineshapes and the relative heights of the signal predicted by the simulations. The simulated Doppler spectra at the four selected lens voltages are plotted, along with the measurements, in Fig.\,\ref{fig:dopplerscan}. The simulation curves shown are a convolution between the lineshapes predicted by trajectory simulation and that of the probe laser, which is stabilized to typically $1\,$MHz. Only one free parameter --- a global vertical scaling on the simulation curves --- is used for all four plots in Fig.\,\ref{fig:dopplerscan}\,(a) to (d). The good agreement in both the lineshapes and relative signal heights across the different lens voltage demonstrates that the numerical model captures very well the dynamics of molecular beam focusing.

It is useful to compare the transverse Doppler width with that obtained in ACME II, when the electric focusing was not available.  This ratio of widths is important to understand how much additional laser power will be needed, relative to before, to saturate the probe signal across the entire velocity distribution when the lens is in operation. In ACME II using a probe laser at $703\,$nm, the $1\,\sigma$ Doppler width was $2.1\,$MHz~\cite{PandaThesis2018}.
The $1\,\sigma$ Doppler width here at the optimal beam focusing (Fig.\,\ref{fig:dopplerscan}\,(c)) is $2.3\,$MHz.  After deconvolving the laser linewidth and accounting for the different probe wavelength here, this is equivalent to a $1\,\sigma$ width of $2.4\,$MHz at $703\,$nm. Thus, at optimized lens voltages, we will need to take into account the $\approx15\%$ wider transverse velocity distribution when preparing and probing the spin precession in ACME III.

\subsection{\label{subsec:spatial}Transverse spatial distribution}
\begin{figure}
	\centering
		\includegraphics[width=0.65\textwidth]{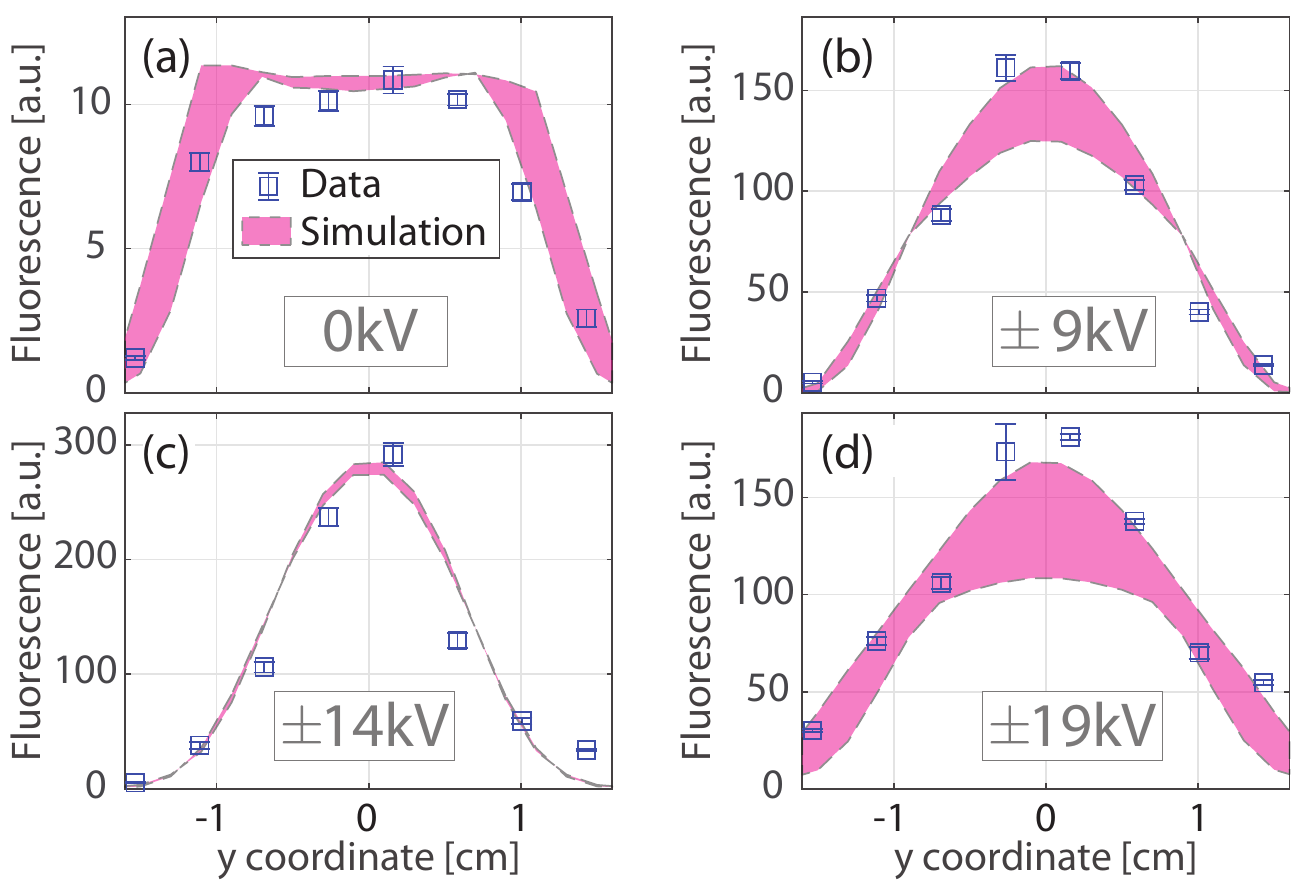}
	\caption{Transverse spatial distributions of the molecular beam along the transverse ($\hat{y}$) direction, at selected lens voltages. Error bars of the data points represent the standard error in the mean over $\approx4$ consecutive measurements, each consisting of $64$ molecule pulses. The range of predictions from trajectory simulations is set by the uncertainty in $\bar{v}_f$ of the molecular beam.}
	\label{fig:Y_scan}
\end{figure}
To characterize the molecular beam size, a spatial scan of the probe laser is performed. Here, the probe laser beam is expanded vertically (along $\hat{y}$) to about $3\,$cm $1/e^2$ diameter, then clipped to $5\,$mm size with a thin slit. This thin slit is scanned along $\hat{y}$ in steps. The fluorescence signal detected at each vertical step, corrected for the spatially dependent photon collection efficiency\footnote{The inhomogeneity in photon collection efficiency results from the fact that the fluorescence collection optics are mounted only above the molecular beam. The functional form of the collection efficiency versus $y$ position is from ray optics simulations, using the known geometry of the collection optics.}, is plotted in Fig.\,\ref{fig:Y_scan} at lens voltages of $0$, $\pm9$, $\pm14$, and $\pm19\,$kV, alongside the predictions from trajectory simulations. The predictions take into account the non-saturated optical transition at the extreme values of the $y$-position  (due to the finite extent of the expanded laser beam), as well as the finite height of the clipped laser beam. Only one global scaling parameter is applied for all $4$ plots of predictions in Fig.\,\ref{fig:Y_scan}. Thus, the experimental data and numerical predictions agree very well for both the spatial distribution and the overall intensity of the molecular beam.

The narrowest spatial profile is obtained when the signal from lens focusing is maximized (at $\pm14\,$kV); here, the FWHM of the distribution is about $1.3\,$cm (after deconvolving the finite height of the probe laser beam). This agrees with the general observation from the sample trajectories in Fig.\,\ref{fig:integrated_signal}\,(b) to (e), where the beam's spatial profile in the detection region is also the narrowest at the condition for the optimal flux. This transverse spatial profile is different from that in ACME II, performed without beam focusing, which had a flat-top distribution (similar to the shape in Fig.\,\ref{fig:Y_scan}\,(a), taken at lens off) with $2.3\,$cm FWHM. This narrower transverse spatial extent of the molecular beam helps both to lower the laser power required to saturate the detection in ACME III, and to in principle reduce the systematic error associated with the transverse gradients of magnetic and electric fields witnessed in ACME II~\cite{ACMECollaboration2018}.

\section{\label{sec:conclusion}Conclusions and outlook}
We have demonstrated a molecular lens system that provides an enhancement of over one order of magnitude in useful flux from a cold beam of heavy ThO molecules, as needed for improving the ACME EDM search. 
We employ a new quantum state preparation scheme that achieves a factor of $1.4$ improvement relative to the rotational cooling scheme in ACME II. This enhanced ground state population is then transferred into the quantum state used for lensing via STIRAP. 
The molecular beam is focused with an electrostatic hexapole lens, yielding a factor of $16$ enhancement in signal size detected from the $Q$ state, compared to when the lens is turned off, in a test beamline similar to the ACME III geometry. \footnote{It is useful to compare this with other configurations, using our validated trajectory simulations. For example, in the case of no lens but a realistic beamline of shorter length connecting the beam source and the expanded EDM measurement region of ACME III, the electrostatic focusing gives a factor of $12$ enhancement in flux. In the case where the EDM measurement region was not expanded but instead remained the same as in ACME II, the molecular lens would provide a factor of $11$ higher flux.} Under these conditions, the transverse velocity and spatial distributions of the lens-focused beam are found to be closely comparable to those used in ACME II. This implies that no significant increase of laser power is required to saturate the preparation and readout of the spin precession, compared to that used in ACME II. 

This flux-enhancing system will be incorporated into the upcoming ACME III measurement. Taking into account the $90\%$ STIRAP efficiency for both pre-lens ($X\textendash C\textendash Q$) and post-lens ($Q\textendash C\textendash X$) state transfer, the molecular beam focusing and improved rotational cooling demonstrated here will enhance the signal size by a factor of $\approx19$ in ACME III.  If shot-noise limited detection can again be achieved such that the sensitivity grows as the square root of the flux, this will correspond to an improvement in the EDM statistical sensitivity by $4.4$ times. This improvement consists of two parts: a factor of $3.5$ from the rotational cooling plus lens system if it was optimized for the shorter beamline in ACME II; and a factor of $1.3$ from additional beam collimation by the lens in the expanded spin-precession region of ACME III. When combined with other demonstrated upgrades, including the longer coherence time to take advantage of the recently-measured $H$ state lifetime ~\cite{Ang2022}, detectors with higher efficiency~\cite{Masuda21}, and a reduction of technical noise~\cite{Panda2019}, we project an improved statistical sensitivity to the electron EDM by potentially $30$ times compared to the current best limit~\cite{ACMECollaboration2018} from ACME II. 

\ack This work was supported by the National Science Foundation, the Gordon and Betty Moore Foundation, and the Alfred P. Sloan Foundation. We thank Oskari Timgren, Vincent Bernado, James Chow, Stan Cotreau, and Jesse Ward for their tremendous technical support. We also thank Zack Lasner, Cristian Panda, Nicholas Hutzler, and Takahiko Masuda for helpful discussions.

\section*{References}

\bibliography{ACMEIIIbib,ACMEIIbib,bib18_fixed} 

\clearpage

\appendix

\section{\label{sec:RotCoolLimit}Data-based limit of rotational cooling gain}
In this section, we derive a data-based upper bound for the population gain in the ground state resulting from the rotational cooling. In an ensemble of ThO molecules in the $X$ electronic state and whose rotational energy levels follow a thermal distribution at temperature $T_{rot}$, the probability to find a particle in the $J$-th rotational energy level is
\begin{equation}
P_J=\frac{1}{Z}(2J+1)\exp\left(-\frac{J(J+1)B_R}{k_BT_{rot}}\right),
\label{eq:boltzmann}
\end{equation}
where $Z=\sum_J{(2J+1)\exp\left(-J(J+1)B_R/(k_BT_{rot})\right)}$ is the partition function, and $B_R=0.33264\,$cm$^{-1}$ is the rotational constant. At $T_{rot}=3.8(5)\,$K~\cite{Hutzler2011a}, the distribution in the lowest few rotational energy levels are given in Table\,\ref{distribution}.
 \begin{table}
\centering
\caption{\label{distribution}Thermal rotational distribution of the $X$ state of ThO at $T_{rot}=3.8\,$K.}
\begin{indented}
\item[]
\begin{tabular}{| l | c| c| c| c| c| c |} \hline \hline \hline
{\bf Rotational level $J$} & $0$ & $1$ & $2$ & $3$ & $4$ & $\geq5$\\ \hline \hline \hline
{\bf Relative population $P_J$ [$\%$] } & \bf{12} & \bf{28} & \bf{28} & \bf{19} & \bf{8.8} & \bf{4.0}\\
\hline \hline\hline
\end{tabular}
\end{indented}
\end{table}

We have previously determined the vibronic branching ratio for the decay $C\,|\nu'=0\rangle\rightsquigarrow X\,|\nu=0\rangle$ in ThO to be $\xi=0.74(6)$~\cite{Wu_2020}, where $\nu$ is the vibrational quantum number. Based on H\"onl-London factors, the rotational branching ratios, $\eta_i$, for $C\,|J'^{\mathcal{P}'}=1^-\rangle\rightsquigarrow X\,|J^\mathcal{P}=i^+\rangle$ (where $i=0$ or $2$) are $\eta_{0}=2/3$ and $\eta_{2}=1/3$. In addition, based on the parity selection rule, spontaneous decay of $C\,|J'^{\mathcal{P}'}=1^-\rangle\rightsquigarrow X\,|J^\mathcal{P}=1^-\rangle$ is forbidden, while decay of $C\,|J'^{\mathcal{P}'}=1^+\rangle\rightsquigarrow X\,|J^\mathcal{P}=1^-\rangle$ has $\eta_{1}=1$. 

When optically pumping on the $P(2)$ line with no external $\mathcal{E}$-field applied, the probability to decay to $X\,|J=i\rangle$ (where $i=0$ or $2$) is $p_{i}=\xi\cdot\eta_{i}$. Thus, pumping on the $P(2)$ transition alone would ideally increase the ground state $X\,|J=0\rangle$ population by 
\begin{eqnarray*}
P_{2,0}&=& p_{0} P_2+p_{0} (P_2 p_{2}) + p_{0} (P_2 p_{2}^2)+ p_{0} (P_2 p_{2}^3)+\cdots\\
            &=& p_{0} P_2 \frac{1}{1-p_{2}}\\
            &=& 18\,\%~(\textrm{of the total rotational population}).
\end{eqnarray*}

The optical pumping on $Q(1)$ takes place with an external field $\mathcal{E}=150\,$V/cm applied, which almost fully mixes the opposite parity levels in $C\,|J'=1\rangle$. This modifies the rotational branching ratios in the decay $C\,|J'=1\rangle\rightsquigarrow X\,|J\rangle$ to be $\eta^*_{J}=1/2\cdot\eta_{J}$. The total branching ratio into $X\,|J\rangle$ is then $p_J^*=\xi\cdot\eta_J^*$. The population that lands in $X\,|J=2\rangle$ will be pumped by the $P(2)$ transition in subsequent cycles. (This is possible because we switch between the $Q(1)$ and $P(2)$ pumping configurations rapidly, and each molecule experiences each laser frequency multiple times during rotational cooling.) Overall, optical pumping on the $Q(1)$ transition (jointly with pumping on the $P(2)$ line) would ideally increase the $X\,|J=0\rangle$ population by
\begin{eqnarray*}
P_{1,0}&=& p_{0}^* P_1+p_{0}^* (P_1 p_{1}^*) + p_{0}^* (P_1 p_{1}^{*2})+ p_{0}^* (P_1 p_{1}^{*3})+\cdots\\
            & &+ p_{0} (P_1 p_{2}^*) + p_{0} (P_1 p_{2}^*) p_2+ p_{0}( P_1 p_{2}^*) p_2^2+ p_{0}( P_1 p_{2}^*) p_2^3+\cdots\\
 	 & &+p_0(P_1 p_1^* p_2^*) + p_0(P_1 p_1^* p_2^*) p_2 +  p_0(P_1 p_1^* p_2^*) p_2^2 +\cdots\\
	 & &+p_0(P_1 p_1^{*2} p_2^*) + p_0(P_1 p_1^{*2} p_2^*) p_2 +  p_0(P_1 p_1^{*2} p_2^*) p_2^2 +\cdots\\
	& &+\cdots\\
	&=& p_{0}^* P_1 \frac{1}{1-p_{1}^*}+p_{0} (P_1 p_{2}^*) \frac{1}{1-p_{2}}\frac{1}{1-p_{1}^*}\\\\
            &=& 14\,\%~(\textrm{of the total rotational population}).
\end{eqnarray*}
Here, the first line takes into account the population falling into $X\,|J=1\rangle$ in the first and subsequent spontaneous decays, and the $n$-th line ($n>1$) accounts for the population landing in $X\,|J=2\rangle$ after the $(n-1)$-th optical pumping cycle, etc. They can all be added in geometric series.

Thus, the expected result for the rotational cooling gain involving initial population in the $J=1$ and $J=2$ levels is $G\equiv(P_0+P_{1,0}+P_{2,0})/P_0=3.7^{+0.4}_{-0.3}$, where the error bars are the quadrature sum of the uncertainties from the vibronic branching ratio, $\Delta G_\xi=^{+0.4}_{-0.3}$, and the rotational temperature in the Boltzmann distribution, $\Delta G_{T}=^{+0.2}_{-0.1}$.

\section{\label{sec:Efield}Electrode configuration between STIRAP region and lens electrodes}
\begin{figure}
	\centering
		\includegraphics[width=0.85\textwidth]{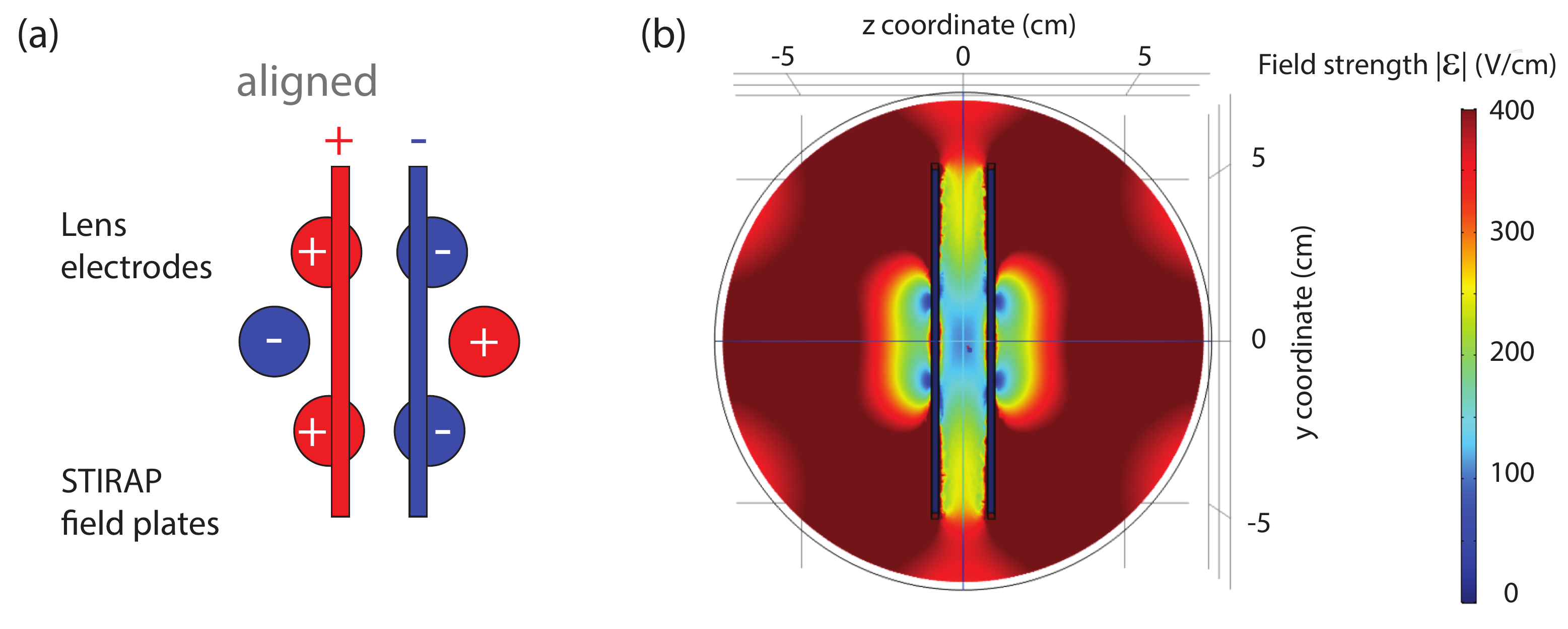}
	\caption{Two arrangements of STIRAP field plates with respect to the hexapole lens electrodes. \textbf{(a)} Each field plate is aligned with the two hexapole electrodes that have the same polarity. In \textbf{(b)} is the corresponding field strength contour in a cross-section plane in the transit region between field plates and hexapole electrodes.}
	\label{fig:E_config}
\end{figure}

To prevent population loss due to nonadiabatic transitions from $Q\,|JM\Omega=2,2,-2\rangle$ into other $M$ sublevels, we need the $\mathcal{E}$-field strength to be nonzero everywhere along the molecules' pathway from the STIRAP region to the lens region. Our electrode configuration to ensure this is shown in Fig.\,\ref{fig:E_config}\,(a), where the field plates are aligned with the two hexapole electrodes of the same polarities. In this case, the $\mathcal{E}_z$ component always stays in the same direction along the molecules' pathway. The resulting $\mathcal{E}$-field strength distribution in the transit region is plotted in Fig.\,\ref{fig:E_config}\,(b). 

\end{document}